\documentclass[11pt]{article}
\usepackage{achemso}
\usepackage{epsfig}

\newcommand{\degree}{$^\circ$}
\newcommand{\beq}{\begin{equation}}
\newcommand{\eeq}{\end{equation}}

\begin{document}

\title{Defects Can Increase the Melting Temperature of DNA-Nanoparticle
Assemblies}

\author{Nolan C.\ Harris and Ching-Hwa Kiang \\ \\
Department of Physics and Astronomy \\
Rice University, Houston, TX\ \ 77005}

\date{
{\em J.\ Phys.\ Chem.\ B,} {\bf 110}, (2006) 16393--16396}

\maketitle

\begin{abstract}

DNA-gold nanoparticle assemblies have shown promise as an alternative
technology to DNA microarrays for DNA detection and RNA profiling.
Understanding the effect of DNA sequences on the melting temperature
of the system is central to developing reliable detection technology.
We studied the effects of DNA base-pairing defects, such as mismatches
and deletions, on the melting temperature
of DNA-nanoparticle assemblies. We found that, contrary to the general
assumption that defects lower the melting temperature of DNA, some
defects increase the melting temperature of DNA-linked nanoparticle
assemblies. The effects of mismatches and deletions were found to
depend on the specific base pair, the sequence, and the location 
of the defects.
Our results demonstrate that the surface-bound DNA exhibit
hybridization behavior different from that of free DNA. Such findings
indicate that a detailed understanding of DNA-nanoparticle assembly
phase behavior is required for quantitative interpretation of
DNA-nanoparticle aggregation.

\end{abstract}

\subsection*{Introduction}

DNA-capped nanoparticle solutions, which self-assemble to form
disordered aggregates, have been shown to exhibit interesting phase
behavior
\cite{Mirkin03a,Kiang03a,Stroud03b,Frenkel04a,Kiang05c,Kiang05b}. In
these systems, the cluster networks are held together by non-covalent
interactions, therefore, the aggregation process is reversible. Unlike
free DNA duplexes, which show a broad transition from double- to
single-stranded DNA, the DNA-nanoparticle assemblies formed here
exhibit a sharp transition from aggregated to dispersed phase
\cite{Kiang03a,Frenkel04a,Kiang05b,Kiang05a}, indicating that melting
of the assembly is not simply a DNA duplex melting process. In
addition, these surface-bound DNA exhibit unusual phase behavior that
deviates from that of the free DNA.

Due to the color change induced by aggregation, DNA-nanoparticle
assemblies have been proposed for use in DNA detection in medical
research, diagnosis of genetic disease, and biodefense
\cite{Whitten03a,Hill00a,Janda02a,Lu04a}, as well as an alternative technology
to DNA microarrays (genechips) \cite{Lipshutz99a} and single-molecule
sequencing \cite{Austin97f}. This nanoparticle technology relies on
differentiation in DNA
hybridization efficiency. In this system, single strands of DNA are
functionalized with an alkanethiol group to bind with gold
nanoparticles. Introducing a specific linker DNA results in aggregation
and a visible color change. The aggregation and melting of these
assemblies are influenced by many parameters, including nanoparticle
size \cite{Kiang03a}, DNA sequence \cite{Kiang05c,Kiang05a} and length
\cite{Mirkin98a,Kiang05b,Mirkin03a}, interparticle distance
\cite{Mirkin98a,Kiang05a}, and electrolyte concentration
\cite{Mirkin03a}.

\begin{figure}[!b]
\begin{center}
\epsfig{file=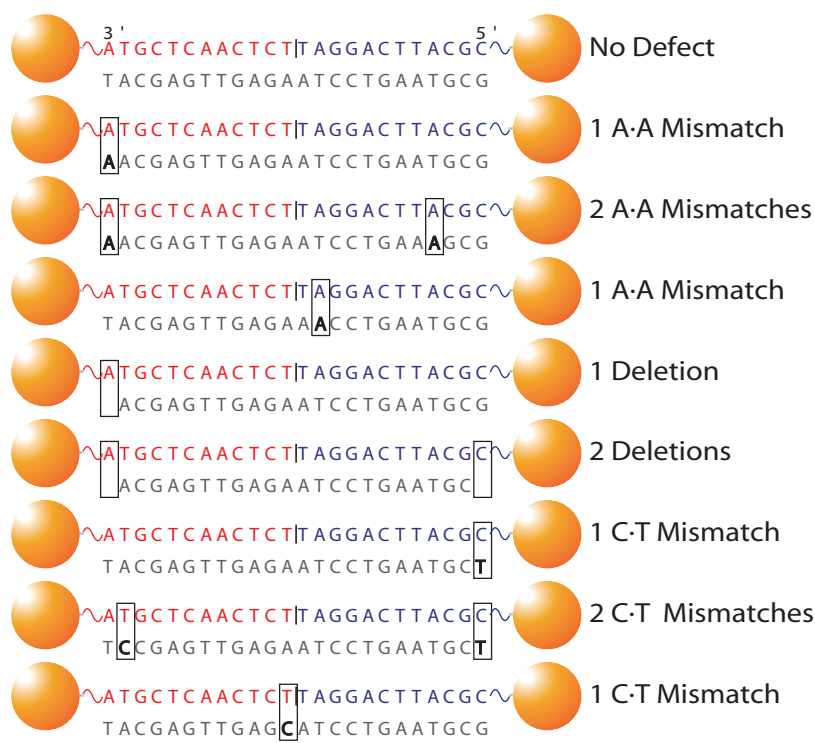,clip=,width=.85\columnwidth}
\end{center}
\caption{Linker and probe DNA sequences used in this study.
Boxed areas indicate base-pairing defects.} \label{fig:fig1}
\end{figure}

Furthermore, self-assembly of DNA-capped gold nanoparticles
has potential to be used for detecting single-base defects
\cite{Mirkin00b,Mirkin98a}. These experiments showed that certain
single-base defects, such as a one base mismatch or deletion, result in
DNA-nanoparticle assemblies with lower melting temperatures ($T_m$)
than assemblies formed using a fully complementary linker. Thus, by
heating DNA-nanoparticle aggregates formed with various linkers to just
below the $T_m$ of their fully complementary counterpart, it is
possible to differentiate solutions containing a complementary 
linker (target) from those with a single-base mismatch. Theory also suggests
that it is possible to detect multiple targets in one solution by
examining the phase behavior of the system
\cite{Frenkel04a,Frenkel05a,Stroud03b}. These technologies assume that
introducing defects results in assemblies with lower $T_m$ than their
fully complementary counterparts \cite{Singh01a,Wattis01a}. While this
assumption holds true for free DNA, it is not directly applicable to
surface-bound DNA. An anomaly in $T_m$ trend has been observed in the
DNA-nanoparticle system \cite{Kiang05c}, which indicates that the details of
DNA base pairing play an important role in the phase behavior of these
nanoparticle systems. Therefore, it is crucial to understand exactly
how the microscopic binding behavior of DNA sequences is mapped onto
the macroscopic phase behavior of DNA-nanoparticle solutions for proper
quantification of data. Here we report experimental observations of
unusual phase behavior in the DNA-nanoparticle system.
Sequence-dependent defects, such as base-pair mismatches and deletions,
were introduced, and $T_m$ trend of the assemblies different from that
of the free DNA has been frequently observed.

\subsection*{Experimental Section}

DNA-capped gold nanoparticles were synthesized and analyzed using
methods described in Refs.[2,5]\cite{Kiang03a,Kiang05c}. Briefly, two
noncomplementary, single-stranded DNA were functionalized with
alkanethiol groups at their ends, to be used as probes.  Probe DNA
were purified by HPLC (Invitrogen) and prepared in 0.3 M NaCl, 0.01 M
phosphate buffer (pH 7). DNA-nanoparticle probes are synthesized by
saturating the surface of colloidal gold particles (Sigma), 10 nm in
diameter, with functionalized probe DNA. Salt was filtered using NAP-5
or NAP-10 columns, to prevent the colloidal gold particles from
irreversible aggregation. 24 hours after mixing the gold nanoparticles
and DNA probes, the solution was centrifuged at 13,200 rpm to
remove excess DNA. Approximately 8 $\mu$l of linker DNA solution
(7$\times$10$^{-6}$ M in 0.3 M NaCl, 0.01 M PBS (pH 7)) was added to
400 $\mu$l of mixed probe solution (4$\times$10$^{17}$ particles/l) and
allowed to aggregate for several days at 4 \degree C.  The
concentration was chosen such that the $T_m$ is independent of linker
concentration.

Nanoparticle assemblies were formed using linkers with either perfectly
matched sequences or with various defects such as mismatches and
deletions. The base pairs found in the usual double-stranded DNA are
the Watson-Crick base pairs (A$\cdot$T and C$\cdot$G), because their
geometry allows any sequence of base pairs to fit into a nucleic acid
sequence without distortion \cite{Crothers00a}. Defects in the current
study include i) mismatched base pairs, and ii) deletions, on or near
the surface, or near the mid-point between two particles (see Fig.
\ref{fig:fig1}). Melting of corresponding sequences of free DNA, which
are not attached to gold nanoparticles, were measured for comparison.
Melting of DNA-nanoparticle aggregates is observed using optical
absorption spectroscopy at 260 nm while solutions are heated at a
constant rate of 1 \degree C/min.
\begin{figure}[!t]
\begin{center}
\epsfig{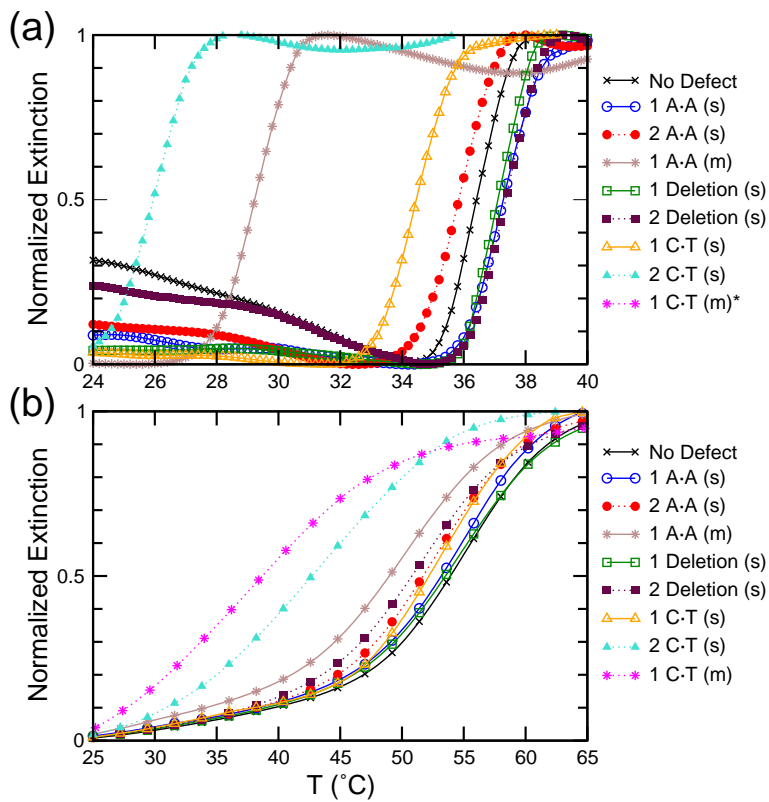}
\end{center}
\caption{Melting curves at 260~nm for (a)
DNA-nanoparticle assemblies and (b) free, unattached DNA
duplexes, formed using linkers with defects of various
composition, number, and location. The ``s'' and ``m'' denote
location of the defect on the surface and middle of the linker,
respectively.} \label{fig:fig2}
\end{figure}

\subsection*{Results and Discussion}

Representative melting curves for nanoparticle assemblies using linkers
with specific defects are shown in Fig. \ref{fig:fig2}(a). Melting
curves for corresponding free DNA duplexes are given by Fig.
\ref{fig:fig2}(b). We found that, unlike the effect of defects on
the free DNA, where mismatches and deletions always lower the $T_m$,
some defects increase $T_m$ in the DNA-nanoparticle assemblies.  For
example, an A$\cdot$A mismatch on the surface has a $T_m$ of 36.1
\degree C, which is higher then the perfectly complementary $T_m$ of
35.2 \degree C, while the corresponding mismatch in a free DNA lowers
$T_m$ from 55.0 \degree C to 54.3 \degree C (see Table
\ref{table:table1}).
\begin{table}[!b]
\vspace{-3ex}
\footnotesize
\begin{center}
\caption{Melting temperatures of DNA-nanoparticle assemblies
and corresponding free DNA duplexes with various base-pairing defects.}
\label{table:table1}
\begin{tabular}{lrrrrr}
\hline
& \multicolumn{2}{c}{$T_m$ (\degree C)} &
\multicolumn{3}{c}{$\Delta T_m$ (\degree C)$^{\mbox{a}}$}   \\ \cline{2-6}
\multicolumn{1}{c}{Defect Type} & Free$^{\mbox{b}}$  &
Bound$^{\mbox{c}}$ & Free$^{\mbox{b}}$ & Bound$^{\mbox{c}}$
& Deviation$^{\mbox{d}}$  \\ \hline
No Defect             & 55.0  & 35.2 & $-$     & $-$      & $-$             \\
1 A$\cdot$A (s$^{\mbox{e}}$)  & 54.3 & 36.1    & $-$0.7   & 0.9    & 1.6    \\
2 A$\cdot$A           & 52.1  & 34.7 & $-$2.9  & $-$0.5   & 2.4             \\
1 A$\cdot$A (m$^{\mbox{f}}$)  & 49.9 & 28.3    & $-$5.1   & $-$6.9 & $-$1.8 \\
1 Deletion  (s)       & 54.4  & 35.9 & $-$0.6  & 0.7      & 1.3             \\
2 Deletion  (s)       & 51.2  & 36.4 & $-$3.8  & 1.2      & 5.0             \\
1 C$\cdot$T (s)       & 53.3  & 33.3 & $-$1.7  & $-$1.9   & $-$0.2          \\
2 C$\cdot$T           & 44.4  & 25.1 & $-$10.6 & $-$10.1  & 0.5             \\
1 C$\cdot$T (m)       & 38.9  & 7.4  & $-$16.1 & $-$27.8  & $-$11.7         \\
\hline
\end{tabular}
\vspace{-2ex}
\end{center}
$^{\mbox{a}}$Change in melting temperature of DNA duplex with defect
compared to corresponding perfectly matched DNA sequence, $\Delta T_m$~=~$T_m$(defect)$-T_m$(no defect). \\
$^{\mbox{b}}$Free DNA. \\
$^{\mbox{c}}$Particle-bound DNA aggregates. \\
$^{\mbox{d}}$Deviation in $\Delta T_m$ of particle-bound DNA versus free DNA. \\
$^{\mbox{e}}$Indicating defects located on surface. \\
$^{\mbox{f}}$Indicating defects located near midpoint between two particles. \\
\vspace{-3ex}
\end{table}
\normalsize

The unusual trend in $T_m$ may be explained by a crowding effect on the
particle surfaces. Replacing a paired base with a mismatched base
allows flexibility in the dangling base (``A'' in this case) to adjust
its position and form non-specific binding with the particle surface,
which increases the $T_m$ of the system, as observed. In fact, Coulomb
blockage is responsible for much of the deviation in DNA hybridization
thermodynamics on surfaces \cite{Pettitt02a}. To further investigate
this effect, we obtained $T_m$ of a system with a ``T'' base deleted
from the linker sequence while keeping the interparticle distance constant 
(see Fig. \ref{fig:fig1}). We found that such
a deletion results in a $T_m$ change of 0.7 \degree C in the
nanoparticle system versus $-$0.6 \degree C for free DNA, which is
consistent with our explanation that deletion of a DNA base at the
end (surface) reduces electrostatic repulsion and, therefore, 
increases the $T_m$.

To examine if these effects are base-dependent, we measured the $T_m$
of a system with one C$\cdot$T mismatch on the surface. Our results
showed that such a defect lowers the $T_m$ ($-$1.9 \degree C) relative
to perfectly complementary particle assemblies, which is similar to the
effect observed in free DNA ($-$1.7 \degree C). This base-dependence
effect (difference between A$\cdot$A and C$\cdot$T) may be understood
in terms of recent experimental results of the binding energy of single
DNA bases on gold surfaces \cite{Mirkin02a,Storhoff02a}. It was
discovered that DNA bases interact with gold surfaces with increasing
strength as T $<$ C $<$ A $<$ G, with the T base interacting much more
weakly than the others. Thus for an A$\cdot$A mismatch near the
particle surface, the non-specific binding between the mismatched A
base and the surface is stronger than that between the T base of
the complementary A$\cdot$T base pair and the surface. On the other
hand, the energy contribution from the binding of a mismatched T base
to the particle surface is known to be much weaker, and thus does not
create more efficient hybridization compared with a complementary
linker. In addition, the energy loss of a disruption of a C$\cdot$G
pair is more than that of an A$\cdot$T pair. Hence, a C$\cdot$G
mismatch will likely result in an overall decreased $T_m$.
It is also possible that, due to the nonspecific binding of the 
end base to the particle surface, the DNA bases near the surface
are partially denatured and do not form base pairs even when the
bases are complementary.  This may explain the small increase in $T_m$
when replacing a complementary base at the surface with a mismatched
base.
 
Sequence dependence effects can also be seen in systems with single base
deletions. This is evidenced by comparing the effects of deleting a T
base from the sequence used here with those seen when deleting a T base
from a much different sequence. It has previously been observed that a
T base deletion may lead to a lowering of $T_m$ \cite{Kiang05c}. In
this study, however, a deletion of a T base results in higher $T_m$.
One difference is that the dangling base after the first deletion is an
A versus a T \cite{Kiang05c}.  As mentioned above, the A base binds to
the particle surface much more strongly than the T base, which may
contribute to the $T_m$ increase in the sequence used here.

\begin{figure}[!b]
\begin{center}
\epsfig{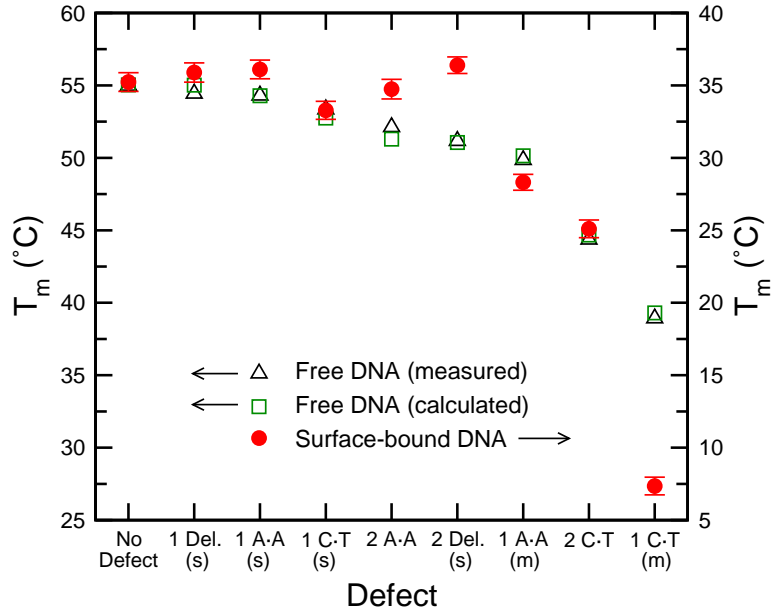}
\end{center}
\caption{Melting temperatures with respect to DNA defect for free,
unattached DNA and surface-bound DNA. Data were taken for 10 nm
DNA-nanoparticle probes and averaged over three separate experiments.
The error bars indicate one standard error. Calculated data represents
predicted $T_m$ values using empirically determined nearest neighbor
thermodynamic parameters for DNA hybridization from 
Refs.[23-26]\cite{Sugimoto96a,SantaLucia99a,SantaLucia98a,SantaLucia00a}.
}
\label{fig:fig3}
\end{figure}

The $T_m$ for free DNA and DNA-nanoparticle assemblies for all defects
are shown in Fig. \ref{fig:fig3}.  $T_m$ for free DNA were calculated
with methods detailed in Ref.[5]\cite{Kiang05c}, using thermodynamic
parameters for base pairs that incorporate nearest neighbor
interactions
\cite{Sugimoto96a,SantaLucia00a,SantaLucia99a,SantaLucia98a}. 
\beq 
{\small
T_m = \left( {\Delta H^0 +~3.4~{{\mbox {kcal} \over \mbox {mol}}} \over
\Delta S^0 - R~\mbox {ln} \left( {1 \over \mbox {[DNA]}} \right)
} \right) + 16.6 \: \mbox {log}_{10} ([\mbox {Na}^+]).
\label{Tm}} 
\eeq
Where $H$ and $S$ are the enthalpy and entropy, respectively, $R$ is the gas
constant, [Na$^+$] and [DNA] are Na$^+$ and DNA concentrations.
Figure
\ref{fig:fig3} shows that the experimentally observed free DNA is well
described by these parameters. The $T_m$ and errors given for the
DNA-nanoparticles system are calculated averages and one standard
errors from repeated experiments. Comparing trends in $T_m$, it is
clear that while $T_m$ decreases with defect for the free DNA system,
it increases for specific defects in the DNA-nanoparticle system.

To test if more than one mismatched or deleted base would contribute to
the detection signal, we prepared systems with two A$\cdot$A or
C$\cdot$T mismatched bases, or two deleted bases (see Fig.
\ref{fig:fig1}). The particle system formed aggregates in the presence
of these defects. Again we found that $T_m$ may be higher or lower
relative to the complementary system, depending on the specific defect,
and that changes in $T_m$ are not always predictable from the free DNA
system (see Fig. \ref{fig:fig3}). Among all defect types studied, the
system with two deletions on the surfaces has the highest $T_m$, as
well as the largest deviation in $T_m$ from its free DNA counterpart.
While two mismatched A$\cdot$A bases near the particle surfaces create
a system with lower $T_m$ compared to the complementary one ($-$0.5
\degree C), the deviation from its free DNA counterpart is significant
(2.4 \degree C). This observation indicates that even two base defects
may contribute significantly to the signal, which should be taken into
account for quantitative analysis of RNA or DNA profiling. The physical
explanation for this finding is similar to that of the one-base defect,
with the magnitude relying on the detailed composition of the bases
involved in the process.

Mismatches at or near the midpoint of the DNA connection, however,
produce very different effects from those near the surface. We found
that an A$\cdot$A or a C$\cdot$T mismatch in the middle of a linker
lowers the $T_m$ for both free and bound DNA.  Since both the free
and surface-bound DNA have the same terminal effect, the stronger effect
of bound-DNA indicates that the effect is amplified
in this system.  For example, a C$\cdot$T mismatch in the middle
significantly lowers the $T_m$ for both free and particle-bound DNA,
with the bound DNA taking several days to form detectable aggregates at
4 \degree C (see Fig. \ref{fig:fig3}). Part of the effect can be
attributed to the fact that the mismatched base does not result in
non-specific binding to the particle surface due to its location, and
thus the only effect is the weakening of the DNA duplex due to the defect.
However, the change in $T_m$ cannot be explained simply by the fact
that there is no surface compensation of the binding energy, since the
second mismatches we introduced in the two mismatch systems have
defects near but not on the surface. The second A$\cdot$A mismatch is
located at the forth base and the second C$\cdot$T mismatch is located
at the second base from the surface. These defects do not result in a
$T_m$ trend similar to those with mismatches in the middle. The
stronger than expected effect of the C$\cdot$T base-pair mismatch in
the middle implies a strong cooperative effect of the particle system
that may stem from the same origin as the asymmetric bond length
disorder of the system \cite{Kiang05c}. We believe that the location of
the defect, which influences the local binding energy distribution, has
an impact on the overall stability of the aggregates.

Similar to our observations for DNA-nanoparticle assemblies,
discrepancy in $T_m$ trends between free and surface-bound DNA has also
been observed in DNA microarrays. DNA microarrays exploit sequence
dependent DNA hybridization in order to quantitatively determine the
level of gene expression in a sample. In some DNA microarrays, for
every DNA probe that is used, a sequence differing by a single base
mismatch is also included in order to determine the amount of
non-specific binding that has occurred \cite{Lipshutz99a}. However,
experimental use of these microarrays has uncovered that, in many
instances, mismatched probes result in more efficient hybridization
than fully complementary probes \cite{Naef02a,Naef03a}. Our results
indicate that when differentiating between a perfectly complementary
linker and those with single- or double-base mismatches, careful
characterization of the behavior of the particle system is required for
quantitative analysis of the results.

The fact that $T_m$ may be higher or lower then the fully complementary
counterpart when there is a base-pairing defect in the DNA sequences
implies that, while DNA-nanoparticle assemblies can distinguish fully
complementary linkers from sequences with defects, quantification of
single-base mismatches or deletions may not be generalized without
detailed characterization of each specific defect. This unusual phase
behavior cannot be predicted by DNA hybridization energy alone, because
surface and cooperative effects influence $T_m$ as well. Both the type
and the location of defects play an important role in the macroscopic
behavior of the system.  Once fully characterized, the results may
be used to increase detection sensitivity by choosing DNA sequences
with defects known to increase $T_m$.

\subsection*{Conclusions}

In conclusion, our results demonstrate that the phase behavior of
DNA-nanoparticle solutions is sensitive to defects in DNA base-paring.
This has implications for the
design of new DNA detection technology, to include DNA-nanoparticle
assemblies and DNA microarrays.
DNA-nanoparticle assemblies remain a promising DNA detection technology
as well as a system with easily controllable parameters for studying
the behaviors of complex fluids.
The complexity of the system should
allow us to probe interesting physics and chemistry that is not
otherwise present in a less controlled system such as a gel.
The system also provides an
opportunity for investigating how a local, microscopic perturbation
affects the macroscopic properties of the system.

\subsection*{Acknowledgments}

We thank the support from NSF DMR-0505814, NIH 1 T90 DK70121-01,
and the Hamill Innovation Fund. \\

\end{document}